\documentclass[preprint]{revtex4-1}
\usepackage{graphicx}
\usepackage{subfig}
\usepackage{color}
\usepackage{amsmath}

\begin{document}

\title{Continuum Random Sequential Adsorption of Polymer on a flat and homogeneous surface.}

\author{Micha\l{} Cie\'sla}
 \email{michal.ciesla@uj.edu.pl}

\affiliation{
M. Smoluchowski Institute of Physics, Jagiellonian University, 30-059 Kraków, Reymonta 4, Poland.}
\date{\today}

\begin{abstract}
Random Sequential Adsorption (RSA) of polymer, modeled as a chain of identical spheres, is systematically studied. In order to control precisely anisotropy and number of degrees of freedom, two different kinds of polymers are used. In the first one, monomers are placed along a straight line; whereas in the second, relative orientations of particles are random. Such polymers fill a flat homogeneous surface randomly. The paper focuses on maximal random coverage ratio and adsorption kinetics dependence on polymer size, shape anisotropy and numbers of degrees of freedom. Obtained results were discussed and compared with other numerical experiments and theoretical predictions.
\end{abstract}

\maketitle

\section{Introduction}
Irreversible adsorption of complex particles at solid and liquid interfaces is of a major significance for many fields such as medicine and material sciences as well as pharmaceutical and cosmetic industries. For example, adsorption of some proteins plays crucial role in blood coagulation, inflammatory response, fouling of contact lenses, plaque formation, ultrafiltration and membrane filtration units operation. Additionally, controlled adsorption is fundamental for efficient chromatographic separation and purification, gel electrophoresis, filtration, as well as performance of bioreactors, biosensing and immunological assays.
\par
Since its introduction by Feder \cite{bib:Feder1980}, Random Sequential Adsorption (RSA) became a well established method of adsorption properties modeling, especially for spherical molecules. On the other hand, using RSA to simulate adsorption of more complicated particles, such as polymers or proteins, raises a question how the universal properties of RSA changes when it is applied to non-spherical molecules. The question has already been answered for basic shapes, e.g. spheroids, spherocylinders, rectangles, needles and similar \cite{bib:Talbot1989, bib:Vigil1989, bib:Tarjus1991, bib:Viot1992, bib:Ricci1992}. However, recent studies shows that such shapes are not sufficient for modeling adsorption of common proteins such as for example fibrinogen \cite{bib:Adamczyk2010}. Therefore, attention of investigators has lately been drawn to to coarse-grained modeling of complex biomolecules and polymers \cite{bib:Adamczyk2011, bib:Rabe2011, bib:Finch2012, bib:Katira2012, bib:Adamczyk2012}.
\par
This study focuses on RSA of polymers on flat and homogeneous two dimensional collector surface. Similar model has been investigated by Adamczyk et al. \cite{bib:Adamczyk2008}; the authors, however, have studied adsorption on squared grid. Other works in this field used different polymers models e.g. \cite{bib:Jia1996} or assumed specific conformations of polymers and used different simulation methods \cite{bib:Sikorski2001}. In all of them, authors have focused on proper modeling of a specific polymer and therefore  numerical method used for adsorption modeling was treated only as a tool. Here the main focus is set on properties of the tool. The main purpose of this work is to find out how the kinetics of RSA as well as basic characteristics of obtained adsorption monolayers depend on particle elongation and its number of degrees of freedom when the particle shape is approximated using coarse-grained approach. In our study, polymer is treated as a kind of a toy-model of complex molecule where both shape anisotropy and number of degrees of freedom are easy to control by merely changing the number of monomers. Therefore the model seems to be the simplest, yet universal tool for determining properties of RSA as well as complex particles adsorption.
\section{Model}
\label{sec:model}
Polymer is modeled as a chain of identical, spherical monomers. In this study, two kinds of polymer models are used:
\begin{description}
\item[i] stiff - monomers are placed along a straight line. In this model, elongation is controlled by number of monomers, whereas number of degrees of freedom remains constant;
\item[ii] flexible - monomer can rotate freely around its neighbors, however, it cannot overlap with other monomers. It is assumed that all monomers lays in a single plane. The number of degrees of freedom increases with monomer count but the increase is non-linear, due to excluded volume effect.
\end{description}
\par
Maximal random coverages are generated using RSA algorithm, which is based on independent, repeated attempts of adding polymer to a covering layer. The numerical procedure is carried out in the following steps:
\begin{description}
\item[i] a virtual polymer is randomly created in such way that the center of each monomer is located on a collector;
\item[ii] an overlapping test is performed for previously adsorbed nearest neighbors of the virtual polymer. The test checks if surface-to-surface distance between monomers is not less than zero;
\item[iii] if there is no overlap the virtual polymer is irreversibly adsorbed and added to an existing covering layer. It's position does not change during further calculations;
\item[iiii] if there is an overlap the virtual polymer is removed and abandoned.
\end{description}
Attempts are repeated iteratively. Their number is commonly expressed in dimensionless time units:
\begin{equation}
t_0 = N\frac{S_P}{S_C},
\end{equation}
where $N$ is a number of attempts, $S_P$ denotes surface area of a single polymer projection on a collector, and $S_C$ is the collector area. Here, $S_P=n\pi r_0^2$ for polymers build of $n$ spherical monomers of radius $r_0$. Square surface of $S_C = (400 \cdot r_0)^2$ with no boundary conditions. Although, in general, specific boundary conditions can influence on obtained results it has been proved that in the case of dimer $(n=2)$ there is no such effect for large enough collectors \cite{bib:Ciesla2012a}. The total number of iterations in each simulation was $10^5 t_0$. Analyzed polymers were build of $2$ to $20$ monomers. For each polymer size and type $100$ independent numerical experiments have been performed.
\section{Results and Discussion}
\begin{figure}[htb]
\vspace{1cm}
\centerline{%
\includegraphics[width=8cm]{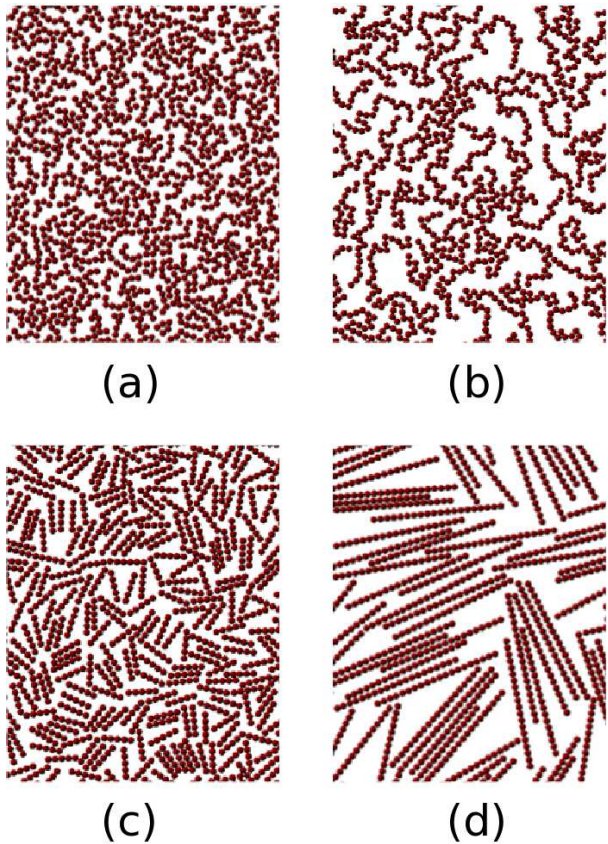}}
\caption{(Color online) Fragments of example layers formed after $t_0 = 10^5$ by flexible (a) (b) and stiff (c) (d) polymer RSA. The polymers are made of $5$ monomers (a) (c) and $20$ monomers (b) (d). Coverage ratios are: (a) - 0.489, (b) - 0.356, (c) - 0.464, (d) - 0.345}
\label{fig:examples}
\end{figure}

Example layers obtained from numerical simulations described in Sec.\ref{sec:model} are presented in Fig.\ref{fig:examples}.
The main parameter characterizing obtained layers is coverage ratio:
\begin{equation}
\theta(t) = n_P(t) \frac{S_P}{S_C},
\end{equation}
which, after infinite iteration of RSA algorithm, approaches the maximal random coverage ratio $\theta_\text{max} = \theta(t \to \infty)$. Parameter $n_P$ denotes here number of adsorbed polymers. To effectively measure $\theta_\text{max}$ using a finite-time computer simulations, appropriate model of adsorption kinetics is needed.
\subsection{RSA kinetics}
\label{sec:kinetics}
Adsorption kinetics can be described analytically for two cases: low coverage limit and jamming limit \cite{bib:Viot1992, bib:Ricci1992}. In general, probability of adsorption depends on uncovered collector area described by Available Surface Function (ASF). For example, when coverage is very low, ASF decreases linearly with a number of adsorbed particles. When coverage increases the ASF decay slows down as two or more particles can block the same space e.g. \cite{bib:AdamczykBook}. Therefore, for a low coverage limit the ASF is often approximated by:
\begin{equation}
B(\theta) = 1 - C_1 \theta + C_2 \theta^2 + o(\theta^2),
\label{eq:asf}
\end{equation}
where $C_1$ corresponds to an area blocked by single molecule and $C_2$ accounts overlapping of those areas \cite{bib:AdamczykBook}. It is worth to notice that $C_1$ and $C_2$ are directly connected with viral coefficients, which can be also calculated from Meyers diagrams, describing adsorbate particles in thermodynamic equilibrium. In case of stiff polymer, the particle shape can be approximated by a spherocylinder. Then, $C_\text{1 (SC)}$ can be analytically derived as \cite{bib:Boublik1975}:
\begin{equation}
C_\text{1 (SC)} = 2\left( 1+\frac{L^2}{4\pi A}\right) = 2\left\{ 1 + \frac{\left[ 2\pi + 4(n-1)\right]^2}{4\pi\left[\pi + 4(n-1)\right]} \right\},
\label{eq:c1sc}
\end{equation}
where $L$ is convex particle perimeter and $A$ is its area. Though spherocylinder  approximates surface blocked by adsorbed polymer particle properly, its area is slightly larger than the polymer model area. Therefore, $C_\text{1 (SC)}$ underestimates the real value and should be multiplied by a ratio of those areas:
\begin{equation}
C_1 = C_\text{1 (SC)} \frac{\pi + 4(n-1)}{n\pi} \approx \frac{8}{\pi}\left(1+\frac{n}{\pi}\right), \mbox{ for large $n$}.
\label{eq:c1}
\end{equation}  
Parameter $C_2$ can be obtained only numerically e.g. \cite{bib:Martinez2006}. In case of the flexible chain, there are no analitical predictions for either $C_1$ or $C_2$ when $n>1$.
\par
\begin{figure}[htb]
\vspace{1cm}
\centerline{%
\includegraphics[width=8cm]{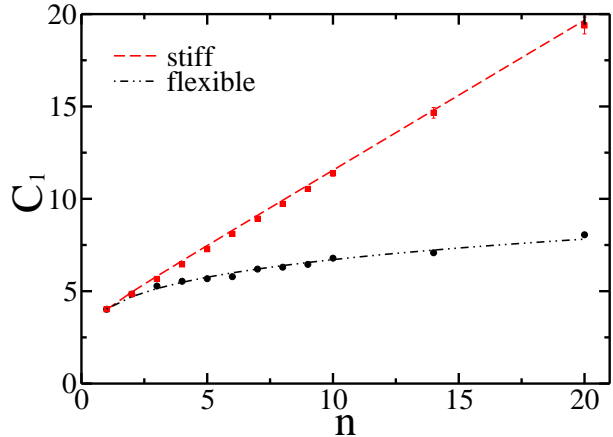}}
\caption{(Color online) Dependence of blocking parameter $C_1$ on polymer size. Dots and squares are simulation data for flexible and stiff polymer, respectively, whereas lines corresponds to fits: (\ref{eq:c1}) for stiff polymer and $C_1 = 4.03 \, n^{0.22}$ for the flexible one.}
\label{fig:c1}
\end{figure}

Parameter $C_1$ has been numerically estimated by fitting $B(\theta)$ defined as (\ref{eq:asf}) for $\theta \le 0.2 \theta_\text{max}$. For stiff polymer, values obtained comply well with the predicted (\ref{eq:c1}). For flexible one, $C_1$ grows with the number of monomers, which can be approximated with a power law (see Fig.\ref{fig:c1}).
As expected, stiff polymer particle blocks more area than the flexible one. Moreover, $C_1$ growth with polymer size is significantly faster for a linear particle.
\par
\begin{figure}[htb]
\vspace{1cm}
\centerline{%
\includegraphics[width=8cm]{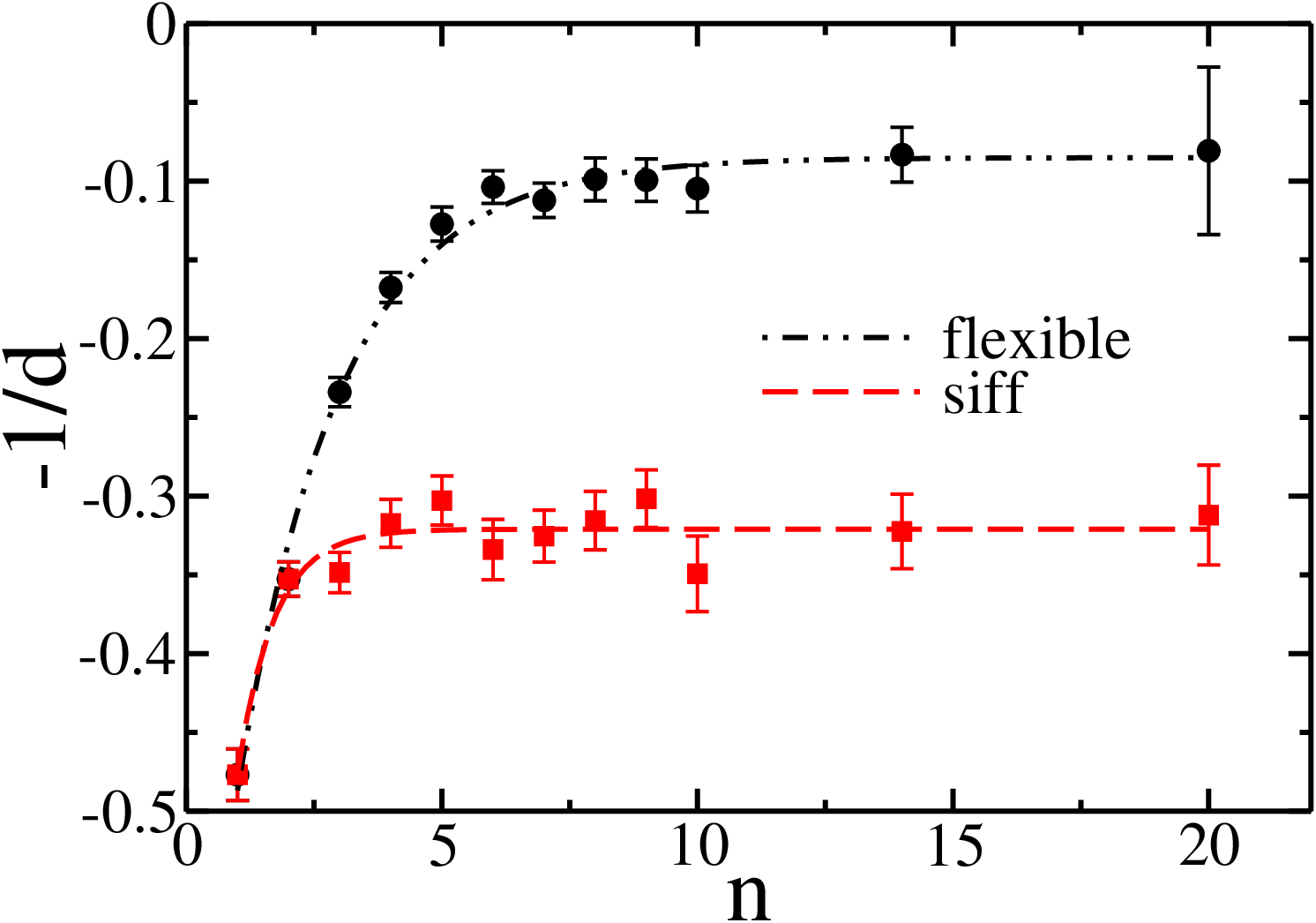}}
\caption{(Color online) Dependence of exponent in (\ref{eq:fl}) on polymer size. Dots and squares are simulation data, whereas lines corresponds to exponential fits: $(-0.32 - 0.54\exp[-1.28 \, n])$ for stiff polymer and $(-0.09 - 0.65\exp[-0.49 \, n])$ for the flexible one.}
\label{fig:alpha}
\end{figure}

The second limit of RSA kinetics (collector almost maximally filled) is of major importance for determining maximal random coverage ratio basing on finite-time simulations. Coverage growth is then governed by Feder's law \cite{bib:Feder1980, bib:Swendsen1981, bib:Privman1991, bib:Hinrichsen1986}:
\begin{equation}
\theta(t) = \theta_\text{max} - A \, t^{-\frac{1}{d}},
\label{eq:fl}
\end{equation}
where $A$ is a coefficient and $d$ is interpreted as collector dimension \cite{bib:Swendsen1981} in case of spherical particles adsorption or more generally as a number of degrees of freedom \cite{bib:Hinrichsen1986}. Feder's law was confirmed for RSA of spheres in several collector dimensions \cite{bib:Torquato2006}, including non-integral  \cite{bib:Ciesla2013c, bib:Ciesla2012b}, as well as for elongated particles \cite{bib:Ricci1992, bib:Ciesla2013a}. The analysis of exponent in (\ref{eq:fl}) estimated using coverage ratio growth presented in Fig.\ref{fig:alpha} reveals at least three things worth  noticing. 
First, the value for $n=1$ (spheres) is close to $-0.5$ as predicted theoretically \cite{bib:Swendsen1981} and confirmed in earlier studies e.g. \cite{bib:Feder1980, bib:Torquato2006}. Second, the value for dimer case ($n=2$) is significantly higher, which reflects more degrees of freedom for such particles. This results differs from  one obtained earlier \cite{bib:Ciesla2012a}. It results from less accurate approximation method used there. Third, the stiff polymer exponent approaches $d\approx 3$ for $n \ge 3$, which reflects third, orientational degree of freedom, and stays at this level despite further increase in monomers number. In the case of flexible polymer, the increase is also curbed, but at higher $d \approx 10$ value. It suggest that hard core interaction between monomers limits an infinite growth of the number of degrees of freedom with polymer length.
\subsection{Maximal random coverage ratio}
\begin{figure}[htb]
\vspace{1cm}
\centerline{%
\includegraphics[width=8cm]{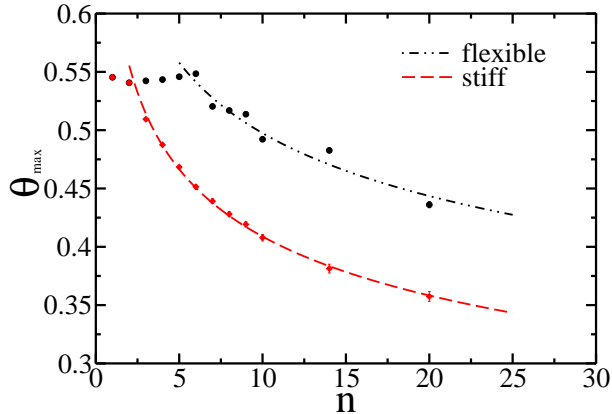}}
\caption{(Color online) Dependence of maximal random coverage ratio on polymer size. Dots and squares are simulation data, whereas lines correspond to power fits: $0.63 \, n^{-0.19}$  for stiff polymer and $0.73 \, n^{-0.17}$ for the flexible one.}
\label{fig:q_n}
\end{figure}

The maximal random coverage ratios were determined by extending obtained RSA kinetics to infinite time. Note that according to the Eq.(\ref{eq:fl}), set of point $(t^{-1/d}, \theta(t))$ obtained from the numerical simulations will form a straight line. By approximating this line up to $t^{-1/d}=0$ one can find $\theta_\text{max}$. Note that the prior knowledge of exponent $-1/d$ estimated in Sec.\ref{sec:kinetics} is essential to get proper values of the maximal random coverage ratio. Results obtained in this way are presented in Fig.\ref{fig:q_n}.
The analysis of RSA of dimer shows, within error margin, the same maximal random coverage ratio for spheres as obtained for dimers \cite{bib:Ciesla2012a}. This, almost constant value of the ratio is observed in the range of $n \le 6$, but only for flexible polymer model. Then, rapid decay of coverage ratio is observed. The existence of the plateau is unexpected. Similar study of RSA on a square lattice shows that the maximal random coverage ratio decreases exponentially with a polymer size \cite{bib:Budinski1997}. However, here on a continuous surface there are at least two competing factors affecting the maximal random coverage ratio. The first is similar as in the lattice case - larger particles are harder to place on a collector due to lower probability of finding a large enough uncovered space. The second factor is a highest packing ratio of monomers in a polymer globule than in a set of independent monomers. The second factor is more important for continuous collectors than for lattice one due to larger possibility of forming a globule, when it is needed. Therefore, in the case of flexible polymer adsorption, competitions of those two factors results in almost constant coverage ratio up to $n \le 6$. There is no such effect for stiff polymer, because there the second factor counts only for $n=2$. The same reason explains much lower (approx. 20\%) values of the maximal random coverage ratio obtained for stiff polymer adsorption. In both the cases, the decay of $\theta_\text{max}$ for large enough polymer length can be approximated by a power or exponential function. To fully discriminate between these two types of relationship the range of studied polymer lengths should be significantly extended.
\subsection{Density autocorrelation and orientational ordering}
\begin{figure}[htb]
\vspace{1cm}
\centerline{%
\includegraphics[width=8cm]{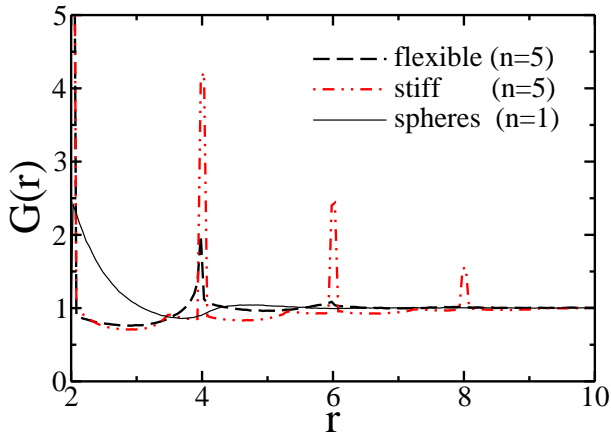}}
\caption{(Color online) Density autocorrelation function. Data for $n=5$ for flexible and stiff polymers were plotted together with data for spheres ($n=1$) used here as a reference level. All functions were obtained for the same coverage ratio $\theta = 0.461$ and were normalized to a mean density of monomers and, therefore, they oscillate around $G(r)=1$.}
\label{fig:g_r}
\end{figure}

Density autocorrelation function $G(r)$, also known as two-point correlation function, is defined here as a mean probability of finding two monomers at distance $r$, regardless of whether they come from the same polymer chain or from different particles. Because the density autocorrelation function depends on the coverage ratio, which is in general different for different polymer length and model, here we decide to calculate density for the coverage ratio close to the maximal one, but equal for all presented cases. Such plot of the $G(r)$ is presented in Fig.\ref{fig:g_r}.
In case of spheres, the density autocorrelation function for maximal random coverages exhibits some universal properties, such as logarithmic singularity for $r \to 2r_0^+$ \cite{bib:Swendsen1981, bib:Privman1991} and fast superexponential decay when $r \gg r_0$ \cite{bib:Bonnier1994}. However, even for relatively short polymers, those properties cannot be observed, due to periodic structure of particles (especially in a stiff polymer). On the other hand, at distances larger than polymer length, almost no density correlation is observed.
\par
\begin{figure}[htb]
\vspace{1cm}
\centerline{%
\includegraphics[width=8cm]{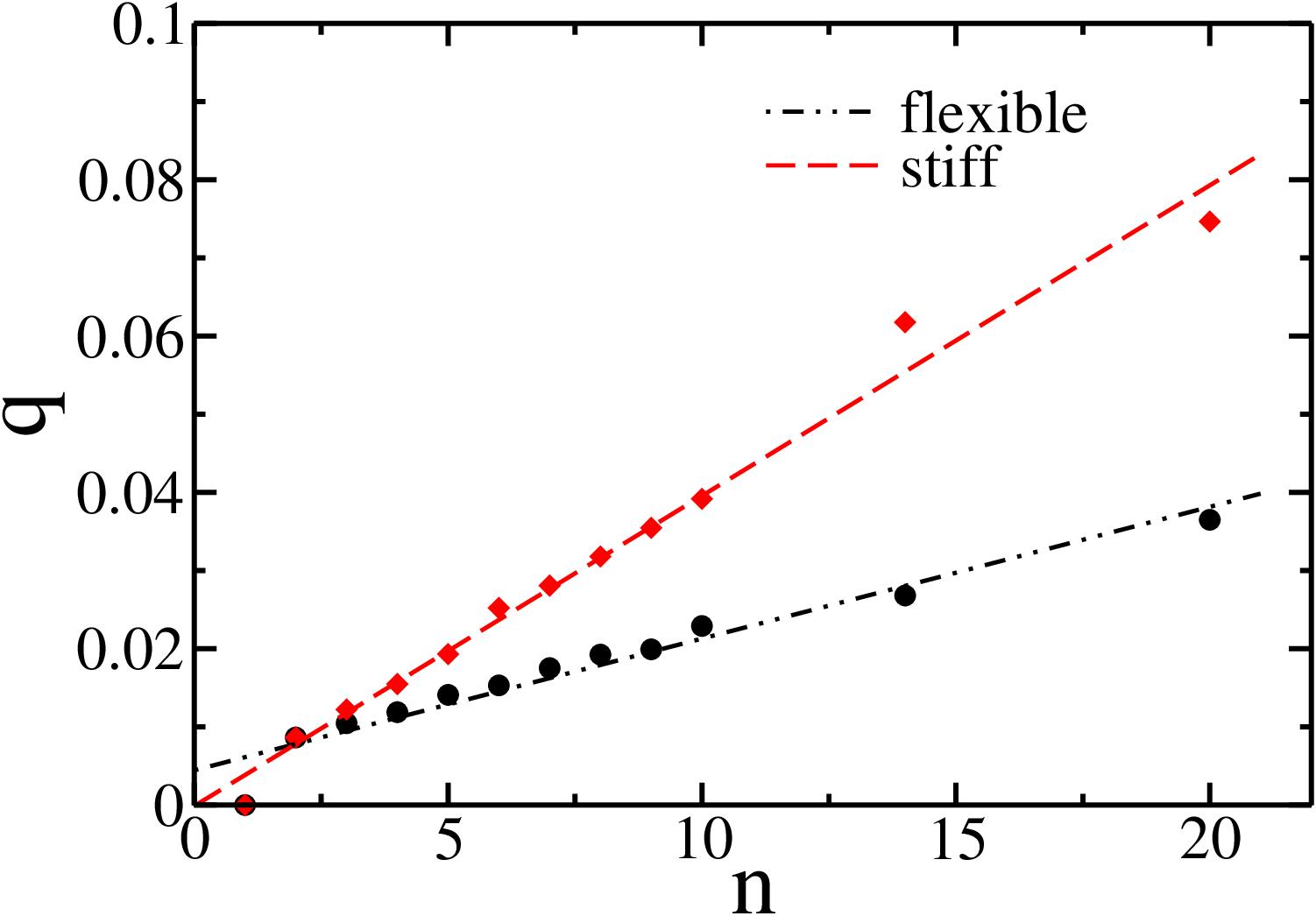}}
\caption{(Color online) Dependence of global orientational order parameter on polymer size. Dots and squares are numerical data, whereas lines are linear fits: $-0.00013 + 0.0040 \cdot n$ for stiff polymer and $y = 0.0044 + 0.0017\cdot n$ for flexible one.}
\label{fig:order_n}
\end{figure}

Elongated particle can form orientationally ordered structures, e.g. liquid crystals. For RSA on infinite collector, when particles orientations are randomly selected according to uniform probability distribution, there is no reason for the global orientational order to appear. Nevertheless, forming of local ordered domains is possible \cite{bib:Ciesla2012a, bib:Ciesla2013c} because parallely aligned particles require less space (see also Fig.\ref{fig:examples}). Therefore, subsequent adsorbed polymers will align parallel to their neighbors, especially when adsorption approaches jamming limit. Situation changes when adsorption on finite collectors is investigated. Adsorption conditions at collector boundaries may promote specific ordering. To measure it quantitatively the order parameter has to be defined \cite{bib:Ciesla2013a}.
\begin{equation}
q = 2 \left[\frac{1}{N} \sum_{i=1}^N (x_i \cos \phi + y_i \sin\phi)^2 - \frac{1}{2} \right],
\end{equation}  
where $N$ is a number of molecules in a layer, $[x_i, y_i]$ is a unit vector pointing from one end of the polymer to the other, and $\phi$ denotes mean direction of all particles and can be calculated as in \cite{bib:Ciesla2012a}. Order parameter $q$ is normalized so as to vanish in totally disordered layers and to equal $1$ for perfectly ordered systems. The mean values of $q$ for obtained coverages are presented in Fig.\ref{fig:order_n}.
As expected, global orientational order increases with polymer size; however, its value is rather small even for the longest stiff polymer because collector area is relatively large. 
\par
To study local ordering, the following function was introduced:
\begin{equation}
q(|\vec{r}|) = 2 \left[ \langle \left[ \phi(\vec{x}) \cdot \phi(\vec{x}+\vec{r}) \right]^2 \rangle -\frac{1}{2} \right], 
\end{equation}
where $\phi(\vec{x})$ is a unit vector along local orientational ordering at point $\vec{x}$ and $\langle \cdot \rangle$ is an average over particle pairs at a distance $r$, measured as a distance between centers of the closest monomers. Relation $q(r)$ shows how the local ordering propagates in a layer.
\begin{figure}[htb]
\vspace{1cm}
\centerline{%
\includegraphics[width=8cm]{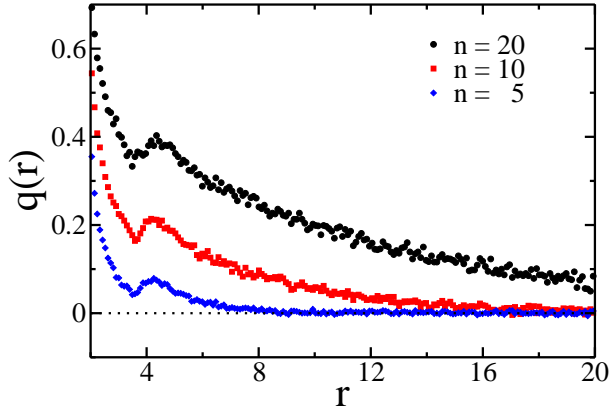}}
\caption{(Color online) Local orientational ordering propagation inside covering layer build of stiff polymers.}
\label{fig:q_r}
\end{figure}
As shown in Fig.\ref{fig:q_r}, ordering vanishes quickly for short stiff polymers. For the longest one, it is significantly larger than $0$ even at distance of $10r_0$, which reflects a tendency for parallel alignment. Slight minimum around $r_0 \approx 3.5$ is connected with the definition of distance between particles, which allows two, even perpendicular polymers, to be as close as parallel ones.
\section{Summary}
In the study, coarse-grain model of polymer is used to test dependence of RSA properties  on number of degrees of freedom and elongation of adsorbate particles. RSA kinetics at low coverage limit is sensitive to molecule shape anisotropy whereas at jamming limit it is governed by number of degrees of freedom, which generally is consistent with Feder's law predictions. Maximal random coverage ratios did not change for short ($n\le 6$) flexible polymers, while rapid drop was observed for stiff polymer starting at $n=3$. Density autocorrelations for maximally covered layers reflect the inner structure of adsorbate particles, especially for stiff polymers. Additionally, in that case, significant local orientational ordering was observed, which reflects domain structure of such layer. 
\par
Flexible and stiff polymer models discussed here, are only two extreme possible cases. In fact, polymer stiffness is controlled by an intra-molecular interactions, which typically depends on an environmental conditions. It is possible to extend presented model of polymer by such interactions and find dependence of properties discussed here on for example temperature in a way as in \cite{bib:Kondrat2002,bib:Kondrat2008}, where similar RSA problem on a square and triangular lattice has been studied.
\par
This work was supported by grant MNiSW/N N204 439040.

\end{document}